# First-principles insights into the mechanical, optoelectronic, thermophysical, and lattice dynamical properties of binary topological semimetal BaGa$_2$

M.I. Naher, S.H. Naqib*
Department of Physics, University of Rajshahi, Rajshahi 6205, Bangladesh
*Corresponding author; Email: salehnaqib@yahoo.com

**Abstract**
In the present study we have investigated the structural properties, electronic band dispersion, elastic constants, acoustic behavior, phonon spectrum, optical properties, and a number of thermophysical parameters of binary topological semimetal BaGa$_2$ in details via first-principles calculations using the density functional theory (DFT) based formalisms. The electronic band structure and density of states calculations with spin orbit coupling reveal semimetallic nature with clear topological signature. The minimum thermal conductivities and anisotropies of the compound are calculated. The elastic constants, phonon dispersion calculations show that the compound under study is both mechanically and dynamically stable. Comprehensive study of elastic constants and moduli shows that BaGa$_2$ possesses fairly isotropic mechanical properties, reasonably good machinability, low Debye temperature and melting point. The chemical bonding in BaG$_2$ is interpreted via the electronic energy density of states, electron density distribution, elastic properties and Mulliken bond population analysis. The compound possesses both ionic and covalent bondings. The reflectivity spectra show strong anisotropy with respect to polarization of the incident electric field in the visible to mid-ultraviolet regions. High reflectivity over wide spectral range makes BaGa$_2$ suitable as a reflecting material. BaGa$_2$ is also an efficient absorber of ultraviolet radiation. Furthermore, the refractive index is quite high in the infrared to visible range. All the energy dependent optical parameters show metallic features and are in complete accord with the underlying bulk electronic density of states calculations. Most of the results presented in this study are novel and should serve as useful reference for future study.

**Keywords:** Topological semimetal; Density functional theory; Elastic properties; Band structure; Optical properties; Thermophysical properties

## 1. Introduction

Since the discovery of superconductivity at 39 K in the layered structure of MgB$_2$ [1], which is also a member of AlB$_2$- type materials, the AlB$_2$-type structures have drawn a lot of attention of condensed matter physicists. In fact, the AlB$_2$-type structures with their superior characteristic features such as the electron–phonon (e–ph) coupling [2], electronic band structure, superconducting state and lattice dynamical properties have resulted in significant amount of theoretical and experimental efforts from the scientific community [2]. BaGa$_2$ is a binary compound in the family of AlB$_2$-type structure. BaGa$_2$ is a superconductor with critical transition temperature $T_c$ = 1.2 K [3]. Furthermore, BaGa$_2$ is a non-magnetic binary compound with electronic band structure exhibiting topological features [2 – 6].

A novel PT (parity-time reversal) symmetry protected Dirac nodal-net semimetal state has been recognized theoretically in AlB$_2$-type structures TiB$_2$ and ZrB$_2$ in the absence of spin orbit coupling (SOC) [4]. Along the same line, BaGa$_2$ has been a recently predicted Dirac semimetal [5, 6]. It was found that BaGa$_2$ is a multilayered Dirac semimetal with quasi-2D Dirac cone located at $E_F$, which is confirmed by the first-principles calculations, angle-dependent dHvA oscillations and ARPES measurements [5]. BaGa$_2$ exhibits unusual interlayer transport properties together with Dirac semimetal behavior under ambient pressure. These features have been explained by the model of tunneling between Dirac fermions in the quantum limit. It should be mentioned that Topological semimetals have attracted tremendous attention of the scientific community due to their exotic electronic properties, such as high charge mobility, large transverse magnetoresistance, non-trivial Berry state, and chiral anomaly [4 – 6]. Beside these exotic features [4 – 6], we have found recently that some of the topological semimetals also possess a number of more conventional physical properties highly pertinent for possible technological applications [7 – 10].

To the best of our knowledge, there are only a few of the physical properties, such as structural, electronic (band, DOS, Fermi surface), phonon spectra, Debye temperature, superconducting transition temperature have been studied for BaGa$_2$ so far [3, 5, 6]. Many other physical properties of the binary BaGa$_2$ are still unexplored. Remarkably, most of the physical properties relevant to potential applications, e.g., charge density distribution, acoustic and thermophysical properties, Mulliken bond population analysis, theoretical hardness, and optical properties of BaGa$_2$ have not been explored at all till date. For instance, analysis of the elastic anisotropy, Cauchy pressure, machinability index, acoustic velocities (both isotropic and anisotropic), acoustic impedance, anisotropy in elastic moduli, and many more are still to be unveiled. A number of thermophysical properties, such as melting temperature, lattice thermal conductivity, thermal expansion coefficient, heat capacity, dominant phonon wavelength and minimum thermal conductivity (both isotropic and anisotropic) have not been investigated thoroughly yet. Besides, the variation in optical constants with incident photon energy is still unknown. In the present study, we aim to explore these unexplored properties of BaGa$_2$ for the first time.

The remaining parts of this manuscript are organized as follows: In Section 2, we have briefly described the methods of *ab-initio* calculations. In Section 3, we have presented the results of our computations and their analyses. Finally, the important features from our study are summarized and discussed in Section 4.

## 2. Computational methodology

The first-principles calculations based on the density functional theory (DFT) [11, 12] are performed using the Cambridge Serial Total Energy Package (CASTEP) code [13]. The generalized gradient approximation (GGA) with the Perdew-Burke-Ernzerhof (PBE) functional [14] is employed to predict the exchange-correlation energy, which depends on both the electron density and its gradient at each point within the crystal. Vanderbilt-type ultra-soft pseudopotential [15] have been used to describe Coulomb potential energy caused by the interaction between the valence electrons and ion cores. It saves us massive computational time with little compromise with computational accuracy. The valence electron configurations for Ba and Ga elements are $5s^2\ 5p^6\ 6s^2$ and $3d^{10}\ 4s^2\ 4p^1$, respectively. Broyden–Fletcher–Goldfarb–Shanno (BFGS) minimization scheme [16] is used for the geometry optimization of the crystal structure of $BaGa_2$. The total energy convergence has been achieved with 500 eV plane-wave cutoff energy. The special $k$-point sampling of the Brillouin zone (BZ) is carried out using the Monkhorst-Pack scheme [17] with a mesh size of 6 × 6 × 5. The geometrical optimization of the $BaGa_2$ crystal volume was performed with the total energy convergence tolerance of $10^{-5}$ eV/atom, maximum lattice point displacement within $10^{-3}$Å, maximum ionic force within 0.03 eVÅ$^{-1}$, maximum stress tolerance 0.05 GPa and energy smearing width 0.1 eV, with finite basis set corrections [18]. The Fermi surface of $BaGa_2$ was obtained by sampling the whole BZ with the $k$-point meshes 38 × 38 × 32. SOC has been incorporated in the electronic band structure calculations to investigate the topological features in the electronic energy dispersion. The phonon dispersion curve (PDC) and related parameters have been calculated using the density functional perturbation theory (DFPT) based finite displacement method (FDM) [19, 20].

Elastic constants of $BaGa_2$ were calculated employing the stress–strain method [21]. Considering the crystal symmetry, a hexagonal system has six independent elastic constants, which are $C_{11}$, $C_{33}$, $C_{44}$, $C_{66}$, $C_{12}$ and $C_{13}$. The calculated values of those independent elastic constants $C_{ij}$ allowed us to evaluate all the macroscopic elastic moduli, such as the bulk modulus ($B$) and shear modulus ($G$) with the Voigte-Reusse-Hill (VRH) approximation [22, 23].

The frequency dependent optical constants of $BaGa_2$ have been calculated from complex dielectric function $\varepsilon(\omega) = \varepsilon_1(\omega) + i\varepsilon_2(\omega)$. The imaginary part of the dielectric function, $\varepsilon_2(\omega)$, related to the photon induced electronic transitions between different electronic states can be obtained from the momentum matrix elements between the occupied and the unoccupied electronic orbitals via the equation:

$$\varepsilon_2(\omega) = \frac{2e^2\pi}{\Omega\varepsilon_0} \sum_{k,v,c} |\langle \Psi_k^c|\hat{u}.\boldsymbol{r}|\Psi_k^v\rangle|^2\ \delta(E_k^c - E_k^v - E) \qquad (1)$$

where, $\Omega$ is the unit cell volume, $\omega$ is the incident light frequency, $\varepsilon_0$ is the dielectric constant of the free space, e is the electric charge, $\hat{u}$ is the unit vector defining the polarization of the incident electric field, $\boldsymbol{r}$ is the position vector, and $\Psi_k^c$ and $\Psi_k^v$ are the conduction and valence band wave functions at a given wave-vector $k$, respectively. The real part of the dielectric function, $\varepsilon_1(\omega)$, has been determined from the corresponding imaginary part $\varepsilon_2(\omega)$ by using the Kramers-Kronig transformation equation. All the other optical constants, namely, refractive index $n(\omega)$, absorption coefficient $\alpha(\omega)$, energy loss-function $L(\omega)$, reflectivity $R(\omega)$, and optical conductivity $\sigma(\omega)$, can be deduced from the estimated values of $\varepsilon_1(\omega)$ and $\varepsilon_2(\omega)$ using standard relations details of which can be found elsewhere [24].

The projection of the plane-wave states onto a linear combination of atomic orbital (LCAO) basis sets [25, 26] has been used for calculating bonding characteristic of BaGa$_2$ employing the Mulliken bond population analysis [27]. Mulliken bond population analysis (MPA) can be carried out using the Mulliken density operator written on the atomic (or quasi-atomic) basis:

$$P_{\mu\nu}^M(g) = \sum_{g'}\sum_{\nu'} P_{\mu\nu'}(g')S_{\nu'\nu}(g-g') = L^{-1}\sum_k e^{-ikg}(P_k S_k)_{\mu\nu'} \qquad (2)$$

and the net charge on atom $A$ is defined as:

$$Q_A = Z_A - \sum_{\mu \in A} P_{\mu\mu}^M(0) \qquad (3)$$

where, the charge on the atomic core is represented by $Z_A$.

As far as the direct calculations of various elastic and thermophysical parameters are concerned, the relevant formulae are given in the respective sections.

### 3. Results and analysis
### 3.1. Structural properties

The crystal structure of BaGa$_2$, a well-known structure of AlB$_2$-type, is hexagonal with space group (*P6/mmm*, No. 191). The unit cell structure of BaGa$_2$ compound is shown in Figure 1. The position of Ba and Ga atoms are at 1a (0, 0, 0) and 2d (1/3, 2/3, 1/2) Wyckoff positions, respectively. The unit cell contains one Ba atom, two Ga atoms. The crystal structure of BaGa$_2$ consists of Ga honeycomb net layers and Ba layers. The estimated lattice parameters of BaGa$_2$ along with available results [3, 28 – 30] are shown in Table 1. The estimated lattice parameters are seen to be in very good agreement with the available experimental data and other results found in the literature.

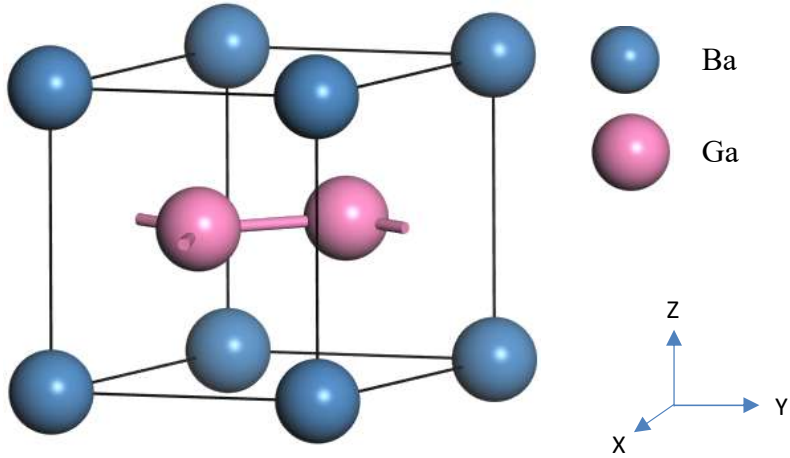

**Figure 1.** Schematic crystal structure of BaGa$_2$. The crystallographic directions are shown.

**Table 1.** Calculated and experimental lattice constants (Å), equilibrium volume $V_0$ (Å$^3$), total number of atoms in the cell of BaGa$_2$.

| Compound | a | b | c | $V_0$ | No. of atoms | Ref. |
|---|---|---|---|---|---|---|
| BaGa$_2$ | 4.463 | 4.463 | 5.124 | 88.40 | 3 | This work |
| | 4.456 | 4.456 | 5.176 | - | - | [3]$^{Theo.}$ |
| | 4.457 | 4.457 | 5.177 | - | - | [28]$^{Expt.}$ |
| | 4.475 | 4.475 | 5.145 | - | - | [29]$^{Expt.}$ |
| | 4.429 | 4.429 | 5.076 | - | - | [30]$^{Expt.}$ |

### 3.2. Mechanical properties

Elastic constants are crucial material parameters. It provides with a link between the mechanical properties and dynamic information concerning the nature of the forces operating in solids, especially for the stability and stiffness of materials. The mechanical properties such as stability, stiffness, brittleness, ductility, and elastic anisotropy of a material are related to the elastic constants which are important for selecting a material for engineering applications. The traditional mechanical stability conditions in hexagonal crystals at equilibrium are expressed in terms of elastic constants as follows: $C_{11} - |C_{12}| > 0$, $(C_{11} + C_{12})C_{33} - 2C_{13}^2 > 0$, $C_{44} > 0$ [31]. The computed elastic constants, listed in Table 2, satisfy the above stability criteria, indicating that BaGa$_2$ is elastically stable.

The Hill approximated values of bulk modulus ($B_H$) and shear modulus ($G_H$) (using the Voigt-Reuss-Hill (VRH) method), Young's modulus ($Y$), Poisson's ratio ($v$) and hardness ($H$) of BaGa$_2$ are estimated with the help of the following relations [32 – 34]:

$$B_H = \frac{B_V + B_R}{2} \tag{4}$$

$$G_H = \frac{G_V + G_R}{2} \tag{5}$$

$$Y = \frac{9BG}{(3B + G)} \tag{6}$$

$$v = \frac{(3B - 2G)}{2(3B + G)} \tag{7}$$

$$H = \frac{(1 - 2v)Y}{6(1 + v)} \tag{8}$$

The shear modulus, $G$ represents the resistance to plastic deformation, while the bulk modulus, $B$ represents their resistance to fracture. For BaGa$_2$, smaller value of $G$ compared to $B$ (Table 3) predicts that the mechanical strength will be limited by the plastic deformation. The bulk modulus depends inversely with the cell volume $V$ ($B \sim V^{-k}$) [35]. Thus the bulk modulus $B$ could be used as a measure of the average atomic bond strength of materials because it has a strong correlation with the cohesion energy or binding energy of the atoms in crystal [36]. The large value of shear modulus, on the other hand, is an indicator of pronounced directional bonding between atoms [37]. The Pugh's ratio, $G/B$, is a widely used parameter which is associated with brittle/ductile nature of a material [38 – 40]. The critical value which separates ductile and brittle materials is around 0.57; i.e., if $G/B > 0.57$, the material behaves in a brittle manner; otherwise the material behaves in a ductile manner. In our case, the $G/B$ value of BaGa$_2$ is 0.54 which indicates that the compound is expected show weakly brittle behavior.

Compressibility, brittle/ductile nature, and characteristic of bonding force of a material are measured by the Poisson's ratio ($v$). The lower and upper limits of $v$ of a material are -1 and 0.5, respectively. For materials with $v = 0.50$, the volume remains unchanged with any amount of deformation within elastic limit. Materials with $v \leq 0.33$ are considered brittle [41]. For central forces, the lower and upper limits of the Poisson's ratio of a material are bounded by 0.25 and 0.50, respectively [42, 43]. The value of the Poisson's ratio for purely covalent materials is of the order of 0.10, whereas for ionic materials a typical value of $v$ is 0.25 [44]. The estimated value of Poisson's ratio of BaGa$_2$ is 0.27, predicting the presence of significant ionic contribution in the bonding, brittle nature and central force in BaGa$_2$. Besides, for ionic and covalent materials, the typical relations between bulk and shear moduli are $G \sim 0.6B$ and $G \sim 1.1B$, respectively [45]. Poisson's ratio is also intimately connected with the way structural elements are packed.

The Cauchy pressure of a compound, an important mechanical parameter for material characterization, is defined as, $(C" = C_{12} - C_{44})$. The value of $C"$ is another approach determining brittle/ductile nature of a material. A positive Cauchy pressure reveals damage tolerance and ductility of a material, while a negative one demonstrates brittleness [46]. The positive Cauchy pressure for BaGa$_2$ indicates that the compound should be ductile in nature. The angular characteristics of atomic bonding in compounds can also be explored with the help of Cauchy pressure. According to Pettifor [47], a material with large positive Cauchy pressure has more metallic non-directional bonding, whereas the Cauchy pressure is negative for directional bonding with angular character. The more negative the Cauchy pressure, more is directional characteristic and lower mobility the bonding. The Cauchy pressure would be zero, if the bonding can be described by pair-wise potentials. The value of $C''$ for BaGa$_2$ is positive, suggesting that the compound under study is ductile with some metallic bonding. Here, we should stress that positive value of the Cauchy pressure of BaGa$_2$ might be somewhat misleading because of the corrections considering many body interaction among atoms and electron gas are not taken into account in determining the elastic constants [48]. The overall behavior of BaGa$_2$ is expected to be weakly brittle; near the borderline between brittleness and ductility.

Machinability is defined as the property of a material which governs the ease or difficulty with which a material can be machined using a cutting tool. This term is widely used in engineering manufacture and production. The machinability depends on many factors related to the work material, cutting tool used, and the cutting parameters. The machinability of a material determines the selection of cutting tool material, geometry of the tool, cutting speed, cutting force, feed rate, and depth of cut. It also determines plasticity [49 – 52] and dry lubricating property of solids. The machinability index, $\mu_M$ of a material can be calculated from [53]:

$$\mu_M = \frac{B}{C_{44}} \tag{9}$$

Thus, high tensile strength combined with low shear resistance leads to good machinability and better dry lubricity. Materials with large value of $B/C_{44}$ possess excellent lubricating properties, lower feed forces, lower friction value, and higher plastic strain value. The $\mu_M$ value of BaGa$_2$ is 1.91, which predicts that the compound possesses good level of machinability comparable to well known MAX phase nanolaminates [54 – 57].

Hardness study of a material is of immense importance for the quality assurance in industry. It explains the load sensitivity of a material. The calculated value of hardness of BaGa$_2$, listed in Table 3, is 3.06 GPa, which is low.

**Table 2.** Calculated elastic constants ($C_{ij}$) (GPa), Cauchy pressure, $C''$ (GPa) of BaGa$_2$.

| Compound | $C_{11}$ | $C_{33}$ | $C_{12}$ | $C_{44}$ | $C_{13}$ | $C_{66}$ | $C''$ | Ref. |
|---|---|---|---|---|---|---|---|---|
| BaGa$_2$ | 66.73 | 57.71 | 34.24 | 22.65 | 18.98 | 16.25 | 11.59 | This work |
|  | 76.6 | 51.8 | 24.6 | 19.4 | 10.9 | 26.0 | - | [3]$^{Theo.}$ |

**Table 3.** The calculated isotropic bulk modulus $B$ (GPa), shear modulus $G$ (GPa), Young's modulus $Y$ (GPa), Pugh's indicator $G/B$, Machinability index $\mu_M$, Poisson's ratio $v$ and hardness $H$ (GPa) of BaGa$_2$ compound deduced from the Voigt-Reuss-Hill (VRH) approximations.

| | B | | | G | | | Y | $\frac{B_V}{B_R}$ | $\frac{G_V}{G_R}$ | G/B | $\mu_M$ | $v$ | H | Ref. |
|---|---|---|---|---|---|---|---|---|---|---|---|---|---|---|
| | $B_V$ | $B_R$ | $B_H$ | $G_V$ | $G_R$ | $G_H$ | | | | | | | | |
| BaGa$_2$ | 37.28 | 36.35 | 36.82 | 20.24 | 19.59 | 19.92 | 50.62 | 1.026 | 1.033 | 0.54 | 1.63 | 0.27 | 3.06 | This work |
| | 33.1 | 31.1 | 32.1 | 23.5 | 22.7 | 23.1 | 55.9 | - | - | 0.72 | - | 0.21 | 5.56 | [3]$^{Theo.}$ |

### 3.3. Elastic anisotropy

Anisotropy in mechanical properties is one of key factor which affect mechanical stability and structural strains of a material under different types of stress. For instance, the creation and propagation of micro-cracks in materials, which is controlled by the anisotropy of mechanical properties, have significant implication in improving the mechanical durability of a material in engineering science. Generally, directional covalent bonding plays a major role in affecting the crystal's anisotropy, while the metallic bonding contributes to improve the overall isotropy.

The shear anisotropic factors can be used to measure degree of anisotropy in atomic bonding in different crystal planes. The shear anisotropy for a hexagonal crystal can be quantified by three different factors [37, 58]:

The shear anisotropy factor for {100} shear planes between the ⟨011⟩ and ⟨010⟩ directions is-

$$A_1 = \frac{4C_{44}}{C_{11} + C_{33} - 2C_{13}} \tag{10}$$

The shear anisotropy factor for the {010} shear plane between ⟨101⟩ and ⟨001⟩ directions is-

$$A_2 = \frac{4C_{55}}{C_{22} + C_{33} - 2C_{23}} \tag{11}$$

and the shear anisotropy factor for the {001} shear planes between ⟨110⟩ and ⟨010⟩ directions is-

$$A_3 = \frac{4C_{66}}{C_{11} + C_{22} - 2C_{12}} \tag{12}$$

The calculated shear anisotropy factors of BaGa$_2$ are listed in Table 5. All three factors equal to 1, when the crystal is isotropic. Otherwise, the crystal is anisotropic. The estimated values of $A_1$ and $A_2$ predict that the compound is moderately anisotropic, but $A_3$ indicates isotropy. $A_1$ is equal to $A_2$, since $C_{11} = C_{22}$, $C_{44} = C_{55}$ and $C_{13} = C_{23}$ for hexagonal structure.

The universal anisotropy index $A^U$, equivalent Zener anisotropy measure $A^{eq}$, anisotropy in shear $A^G$ (or $A^C$) and anisotropy in compressibility $A^B$ of solids with any symmetry can be estimated by the following standard equations [59 – 62]:

$$A^U = 5\frac{G_V}{G_R} + \frac{B_V}{B_R} - 6 \geq 0 \tag{13}$$

$$A^{eq} = \left(1 + \frac{5}{12}A^U\right) + \sqrt{\left(1 + \frac{5}{12}A^U\right)^2 - 1} \tag{14}$$

$$A^G = \frac{G^V - G^R}{2G^H} \tag{15}$$

$$A^B = \frac{B_V - B_R}{B_V + B_R} \tag{16}$$

Universal anisotropy index ($A^U$) is one of the most widely used indices to quantify anisotropy in elastic properties. It is a singular measure of anisotropy irrespective of the crystal symmetry. $A^U$ is the first anisotropy parameter which accounts both shear and bulk contributions, unlike all other existing anisotropy parameters. From Eqn. 13, we can say that $G_V/G_R$ has greater influence on the anisotropy index $A^U$ than $B_V/B_R$. $A^U$ is zero for an isotropic material, whereas a value smaller or greater than zero suggests varying degrees of anisotropy. The larger $A^U$ means the stronger anisotropy. The calculated values of $A^U$ of BaGa$_2$ is 0.19, showing anisotropic properties.

The value of $A^{eq}$ is 1.0 for locally isotropic crystals. The calculated values of $A^{eq}$ for BaGa$_2$ at 0 GPa is 1.49 predicting that the crystal is anisotropic. The values of $A^B$ and $A^G$ ranged from 0 to 1. $A^B = A^G = 0$ and $A^B = A^G = 1$ represent the perfect elastic isotropy and maximum elastic anisotropy, respectively. The larger value of $A^G$ in comparison with $A^B$ (Table 4) indicates that anisotropy in shear is larger than the anisotropy in compressibility.

A log-Euclidean formula is used to define a universal log-Euclidean index as [59, 63]:

$$A^L = \sqrt{\left[\ln\left(\frac{B^V}{B^R}\right)\right]^2 + 5\left[\ln\left(\frac{C_{44}^V}{C_{44}^R}\right)\right]^2} \tag{17}$$

where $C_{44}^V$ and $C_{44}^R$ are the Voigt and Reuss approximated values of $C_{44}$, respectively.

The values of $C_{44}^V$ and $C_{44}^R$ are obtained from [59]:

$$C_{44}^R = \frac{5}{3}\frac{C_{44}(C_{11} - C_{12})}{3(C_{11} - C_{12}) + 4C_{44}} \tag{18}$$

and

$$C_{44}^V = C_{44}^R + \frac{3}{5}\frac{(C_{11} - C_{12} - 2C_{44})^2}{3(C_{11} - C_{12}) + 4C_{44}} \tag{19}$$

Like the universal anisotropy index, the expression for $A^L$ is valid for any crystal symmetry. This index is scaled correctly for perfect isotropy and is valid for all crystallographic point symmetry groups. However, $A^L$ is found to be more appropriate for the present analysis for it is less sparse than $A^U$ when considering extremely anisotropic crystallites. $A^U$ cannot explain the absolute level of anisotropy but merely anisotropic nature. Hence, $A^L$ calculation, using the difference between the averaged stiffnesses $C^V$ and $C^R$, is considered to be more appropriate for anisotropy study. The values of $A^L$ range between 0 to 10.26, and greater than 1 for almost 90% of solids. In case of complete isotropy, $A^L$ is equal to zero. The estimated value of $A^L$ for BaGa$_2$ is 0.24, smaller than 1, indicating moderate anisotropy. Generally, it has been claimed that materials with higher (lower) $A^L$ indicate the presence of layered (non-layered) type structure, respectively [59, 64, 65]. The comparatively low value of $A^L$ implies that BaGa$_2$ does not exhibit layered feature.

The linear compressibility of a hexagonal crystal along $a$ and $c$ axis ($\beta_a$ and $\beta_c$) are evaluated from [66]:

$$\beta_a = \frac{C_{33} - C_{13}}{D} \quad \text{and} \quad \beta_c = \frac{C_{11} + C_{12} - 2C_{13}}{D} \tag{20}$$

with $D = (C_{11} + C_{12})C_{33} - 2(C_{13})^2$

The calculated values are listed in Table 4. Elastically isotropic crystals have unit value for $\beta_c/\beta_a$. The deviation of these factors from unit value quantifies the level of elastic anisotropy in compression. The calculated values indicate that compressibility along $a$ axis and $c$ axis are anisotropic in nature.

**Table 4.** Shear anisotropy factor ($A_1$, $A_2$ and $A_3$), the universal anisotropy index $A^U$, equivalent Zener anisotropy measure $A^{eq}$, anisotropy in shear $A_G$ (or $A^C$) and anisotropy in compressibility $A_B$, universal log-Euclidean index $A^L$, linear compressibility ($\beta_a$ and $\beta_c$) (TPa$^{-1}$) and their ratio $\beta_c/\beta_a$ for BaGa$_2$.

| Compound | $A_1$ | $A_2$ | $A_3$ | $A^U$ | $A^{eq}$ | $A_G$ | $A_B$ | $A^L$ | Layered | $\beta_a$ | $\beta_c$ | $\beta_c/\beta_a$ |
|---|---|---|---|---|---|---|---|---|---|---|---|---|
| BaGa$_2$ | 1.05 | 1.05 | 1 | 0.19 | 1.49 | 0.02 | 0.01 | 0.17 | No | 0.008 | 0.012 | 1.62 |

The bulk modulus of a material also has directional dependence. The bulk modulus of a solid along different crystallographic axes can be calculated either from the pressure dependent lattice parameter measurements or by means of the single crystal elastic constants. It is simpler to calculate the bulk modulus along three axes from the single crystal elastic constants. The relaxed bulk modulus and uniaxial bulk modulus along $a$, $b$ and $c$ axis and anisotropies of the bulk modulus can be defined as follows [58]:

$$\frac{B_{relax}}{= \frac{\Lambda}{(1+\alpha+\beta)^2}} \quad ; \quad \begin{aligned} B_a &= a\frac{dP}{da} \\ &= \frac{\Lambda}{1+\alpha+\beta} \end{aligned} \quad ; \quad \begin{aligned} B_b &= a\frac{dP}{db} \\ &= \frac{B_a}{\alpha} \end{aligned} \quad ; \quad \begin{aligned} B_c &= c\frac{dP}{dc} \\ &= \frac{B_a}{\beta} \end{aligned} \quad (21)$$

and

$$A_{B_a} = \frac{B_a}{B_b} = \alpha \quad ; \quad A_{B_c} = \frac{B_c}{B_b} = \frac{\alpha}{\beta} \quad (22)$$

where,

$$\Lambda = C_{11} + 2C_{12}\alpha + C_{22}\alpha^2 + 2C_{13}\beta + C_{33}\beta^2 + 2C_{23}\alpha\beta$$

$$\alpha = \frac{(C_{11} - C_{12})(C_{33} - C_{13}) - (C_{23} - C_{13})(C_{11} - C_{13})}{(C_{33} - C_{13})(C_{22} - C_{12}) - (C_{13} - C_{23})(C_{12} - C_{23})}$$

and

$$\beta = \frac{(C_{22} - C_{12})(C_{11} - C_{13}) - (C_{11} - C_{12})(C_{23} - C_{12})}{(C_{22} - C_{12})(C_{33} - C_{13}) - (C_{12} - C_{23})(C_{13} - C_{23})}$$

where, $A_{B_a}$ and $A_{B_c}$ are anisotropies of bulk modulus along the $a$ axis and $c$ axis with respect to $b$ axis, respectively.

The calculated values are listed in Table 5. For BaGa$_2$, $A_{B_a} = 1$ and $A_{B_b} \neq 1$, which implies anisotropy in axial bulk modulus. It is observed that bulk modulus along $c$ axis is smaller than those along $a$ and $b$ axis. Generally, the values of uniaxial Bulk modulus are different and significantly larger than the isotropic bulk modulus. This comes from the fact that the pressure in a state of uniaxial strain for a given crystal density generally differs from the pressure in a state of hydrostatic stress at the same density of the solid [37].

**Table 5.** B$_{relax}$, uniaxial bulk moduli (in GPa) and anisotropy indices for bulk modulus along different axes of BaGa$_2$.

| Compound | $B_{relax}$ | $B_a$ | $B_b$ | $B_c$ | $A_{B_a}$ | $A_{B_c}$ |
|---|---|---|---|---|---|---|
| BaGa$_2$ | 36.35 | 131.85 | 131.85 | 81.04 | 1.0 | 0.61 |

To visualize elastic anisotropy in further details, we have estimated the direction dependent variation of Young's modulus, compressibility, shear modulus and Poisson's ratio using the ELATE code [67]. Figure 2 demonstrates the 3D plot of Young's modulus, compressibility, shear modulus and Poisson's ratio. The 3D contour plots must be spherical for isotropic crystal;

otherwise, it demonstrates the degree of anisotropy. As can be seen from Fig. 2, there is small deviation from spherical shape in the 3D figures of *Y*, *K*, *G* and *v* signifying some degree of anisotropy. The ELATE generated plots also demonstrate that the projections of direction dependent *Y*, *K*, *G* and *v* in the *ab*-plane are fairly circular. This implies that the elastic anisotropy within the basal plane is very small.

The maximum and minimum values of *Y*, *K*, *G*, *v* and their maximum to minimum ratios are presented in Table 6. These ratios are useful indicators of elastic anisotropy.

**Table 6.** The minimum and maximum values of Young's modulus (GPa), compressibility (TPa$^1$), shear modulus (GPa), Poisson's ratio and their ratio of BaGa$_2$.

| Phase | Y | | $A_Y$ | K | | $A_K$ | G | | $A_G$ | v | | $A_v$ |
|---|---|---|---|---|---|---|---|---|---|---|---|---|
| | $Y_{min}$ | $Y_{max}$ | | $K_{min}$ | $K_{max}$ | | $G_{min}$ | $G_{max}$ | | $v_{min}$ | $v_{max}$ | |
| BaGa$_2$ | 47.528 | 51.858 | 1.1091 | 7.586 | 12.339 | 1.627 | 16.245 | 22.651 | 1.394 | 0.139 | 0.463 | 3.335 |

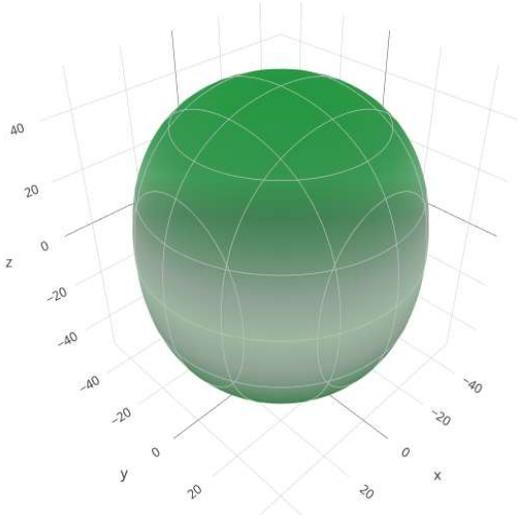

(a)

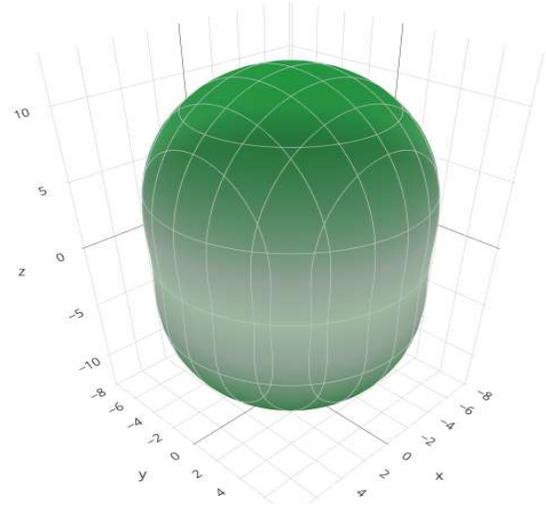

(b)

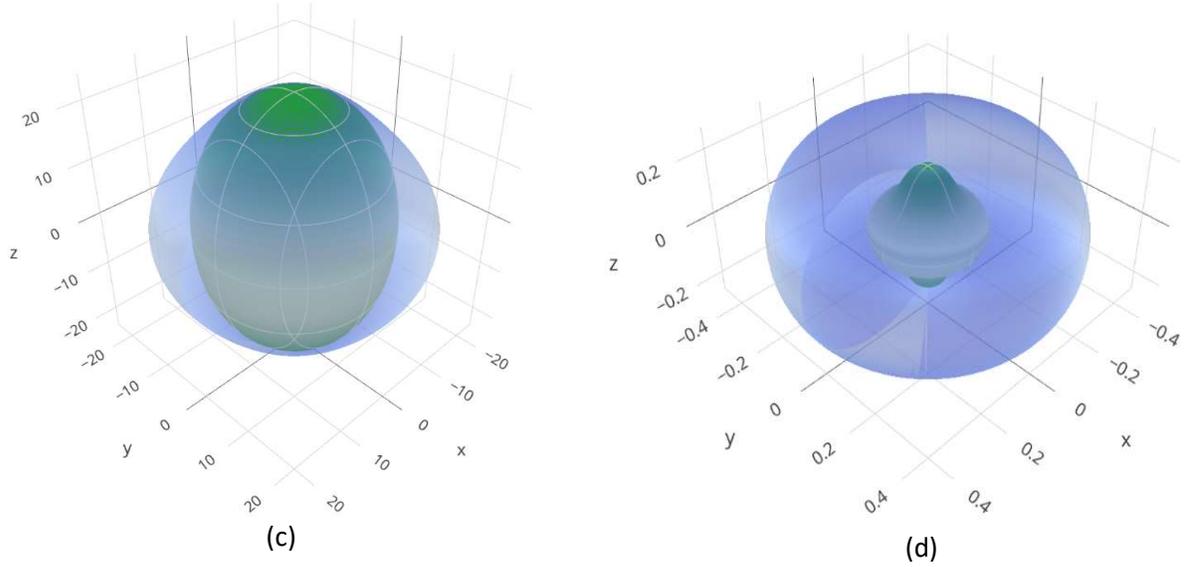

(c)             (d)

**Figure 2.** 3D directional dependences of (a) Young modulus (b) linear compressibility (c) shear modulus and (d) Poisson's ratio for $BaGa_2$ compound.

## 3.4. Acoustic velocities and their anisotropy

The elastic properties of a material are closely linked to the propagation of elastic (sound) velocities through it. The research on elastic waves has been a key interest in physics, materials science, seismology, geology, musical instrument designing, and medical sciences. Moreover, the acoustic behavior of a material provides a link between the thermal and electrical conductivity of a material. According to the classical theory, the sound velocity of a crystal is directly related to its thermal conductivity, $k = \frac{1}{3}C_v l v$. The transverse and longitudinal acoustic waves traversing through a crystalline solid can be calculated with the help of the bulk and shear moduli using the equations [68]:

$$v_t = \sqrt{\frac{G}{\rho}} \quad \text{and} \quad v_l = \sqrt{\frac{B + 4G/3}{\rho}} \quad (23)$$

where, $\rho$ is the mass-density of the material.

The average sound velocity $v_a$ in the crystal is estimated using following equation [68]:

$$v_a = \left[\frac{1}{3}\left(\frac{2}{v_t^3} + \frac{1}{v_l^3}\right)\right]^{-\frac{1}{3}} \quad (24)$$

The estimate acoustic velocities are given in Table 7.

The acoustic impedance is a useful parameter which determines the transfer of acoustic energy between two media. The amount of reflected and transmitted energy, when sound wave arrives at the interface of two media, can also be expressed in terms of their difference in acoustic impedances. Large sound pressure creates a large particle velocity if the impedance of the medium is low, but the same sound pressure will create only a relatively small particle velocity if the acoustic impedance of the medium is high. As a result, most of the sound gets transmitted or reflected, if the impedance difference is about equal or much higher, respectively. The study of acoustic impedance has a vast application in transducer design, acoustic sensor, aircraft engine noise reduction, industrial factories, automobiles and many underwater acoustic applications. The acoustic impedance of a material can be evaluated from [64]:

$$Z = \sqrt{\rho G} \tag{25}$$

Acoustic impedance of a material changes with temperature, since both density and elastic constants depend on temperature. Although the temperature dependence is much weaker in case of shear modulus compared to that for the density.

Every vibrating material radiates acoustic energy and the intensity of this radiation is another important parameter which is used to select materials for sound boards design. The front plate of a violin, the sound board of a harpsichord and the panel of a loudspeaker are examples. The intensity, $I$, is proportional to the surface velocity and defined as [64, 69]:

$$I \approx \sqrt{G/\rho^3} \tag{26}$$

where, $\sqrt{G/\rho^3}$ is the *radiation factor*. Generally, materials with high *radiation factor* are considered to be suitable for sound boards. The most widely used material for the front plate of violins is spruce (8.6 m$^4$/kg.s). On the other hand, maple (5.4 m$^4$/kg.s) is used for the back plate of violins whose function is to reflect, not radiate. The evaluated radiation factor for BaGa$_2$ is listed in Table 7.

We have also calculated the Grüneisen constant $\gamma$ of BaGa$_2$ compound is which is calculated from the Poisson's ratio by using the following expression [70]:

$$\gamma = \frac{3(1+\nu)}{2(2-3\nu)} \tag{27}$$

This parameter controls number of important physical processes, like thermal conductivity, thermal expansion, absorption of acoustic waves and the temperature dependence of elastic properties. It also gives a measure of the anharmonic effect, i.e., the temperature dependence of phonon frequencies and phonon damping as well as the thermal expansion effects. The calculated value of $\gamma$ of BaGa$_2$ is 1.60, tabulated in Table 7. This value is quite typical of crystalline solids [71].

**Table 7.** Density $\rho$ (g/cm$^3$), transverse velocity $v_t$ (ms$^{-1}$), longitudinal velocity $v_l$ (ms$^{-1}$), average elastic wave velocity $v_a$ (ms$^{-1}$), Grüneisen parameter $\gamma$, acoustic impedance $Z$ (Rayl) and radiation factor $\sqrt{G/\rho^3}$ (m$^4$/kg.s) of BaGa$_2$.

| Compound | $\rho$ | $v_t$ | $v_l$ | $v_a$ | $\gamma$ | $Z$ | $\sqrt{G/\rho^3}$ |
|---|---|---|---|---|---|---|---|
| BaGa$_2$ | 5.20 | 1957.39 | 3491.53 | 2161.80 | 1.60 | 10.18 | 0.377 |

In solids each atom has three modes of vibrations, one longitudinal and two transverse modes. In an anisotropic compound the pure longitudinal and transverse modes can exist only along certain directions. On the other hand, the propagating modes in all other directions are either quasi-transverse or quasi-longitudinal. The hexagonal symmetry of the compound dictates that pure transverse and longitudinal modes can only exist for the symmetry directions along [100] and [001]. The acoustic velocities of $BaGa_2$ in the principal directions can be calculated from the single crystal elastic constants [72]:

[100]:

$$[100]v_l = \sqrt{(C_{11} - C_{12})/2\rho}; \quad [010]v_{t_1} = \sqrt{C_{11}/\rho}; \quad [001]v_{t_2} = \sqrt{C_{44}/\rho}$$

[001]:
(28)

$$[001]v_l = \sqrt{C_{33}/\rho}; \quad [100]v_{t_1} = [010]v_{t_2} = \sqrt{C_{44}/\rho}$$

where $v_{t_1}$ and $v_{t_2}$ are the first and second transverse modes, respectively. $v_l$ is the longitudinal sound velocity, $\rho$ is the density. The velocity of elastic waves inside a material depends only on its (material's) nature, neither on its dimension nor on its (wave's) frequency. The calculated sound velocities in the directions are tabulated in Table 8. The longitudinal velocity along [001] is larger than along [100].

**Table 8.** Anisotropic sound velocities (ms$^{-1}$) of $BaGa_2$ along different crystallographic directions.

| Propagation directions | | Sound velocity |
|---|---|---|
| [100] | $[100]v_l$ | 1767.66 |
| | $[010]v_{t_1}$ | 3582.62 |
| | $[001]v_{t_2}$ | 2087.25 |
| [001] | $[001]v_l$ | 3331.70 |
| | $[100]v_{t_1}$ | 2087.25 |
| | $[010]v_{t_2}$ | 2087.25 |

### 3.5. Electronic properties
#### 3.5.1. Electronic band structure

The electronic band structure calculation is helpful to understand many physical properties of crystalline solids. Band structure almost completely explains optical as well as charge transport properties. We have calculated the electronic band structure for $BaGa_2$ along high symmetry directions within the $k$-space as a function of energy ($E$-$E_F$), which is displayed in Fig. 3. The Fermi level $E_F$ is shown as horizontal broken line (Fig. 3). The total number of bands is 42. The band structure indicates the metallic nature of BaGa$_2$ as there is a noticeable overlapping of conduction bands and valence bands at the Fermi level. The bands crossing the Fermi level are shown in different colors with their corresponding band numbers. The extent of band overlap and crossings of the Fermi level are not strong. The bands below the Fermi level mainly originated from the Ba-4$d$ and Ga-4$p$ electronic orbitals. The conduction band is also formed from the Ba-4$d$ and Ga-4$p$ electronic states. The computed band structure shows excellent agreement with prior results [5, 6]. BaGa$_2$ exhibits clear semimetallic character with Dirac cone located at the $E_F$. The band dispersions reveal anisotropy. Electronic dispersions running along the $c$-direction are less dispersive compared those in the $ab$-plane. Therefore, the effective mass of electrons and various charge transport coefficients are expected to show direction dependence. The quasi-Dirac cone located at the $K$-point of the BZ at Fermi energy indicates that the mobility of electrons residing in these bands should be very high.

(a)  (b)

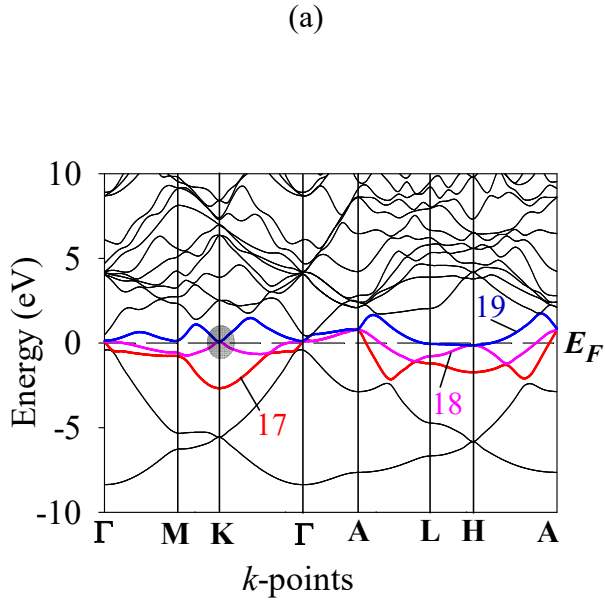
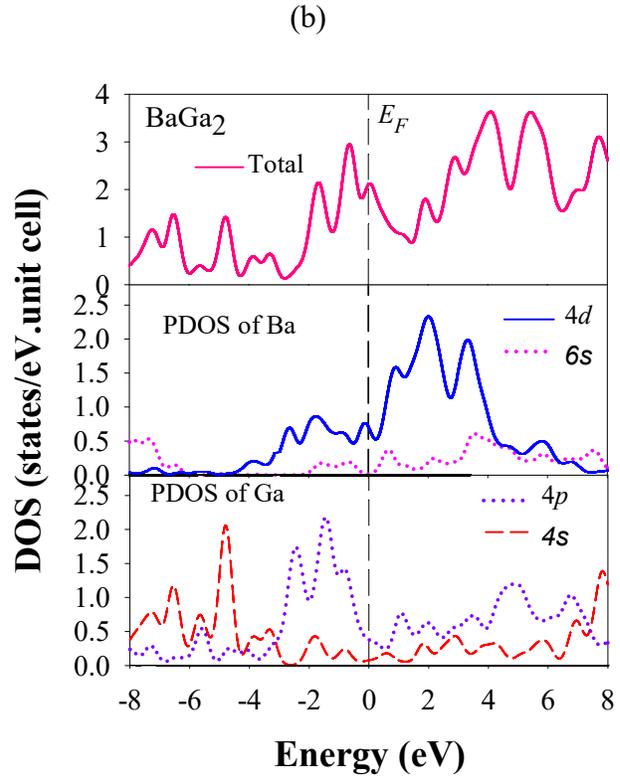

**Figure 3.** Electronic band structure of BaGa$_2$ with SOC along several high symmetry directions in the Brillouin zone. The gray circle locates the Dirac cone confirming the topological character.

**Figure 4.** Total and partial electronic density of states (PDOSs and TDOSs) of BaGa$_2$ with SOC as a function of energy. The Fermi level is placed at zero energy.

### 3.5.2. Density of states

To further clarify the contributions of different orbitals to the electronic properties and nature of chemical bonding, total density of states (TDOS) and partial density of states (PDOS) of BaGa$_2$ have been calculated and displayed in Fig. 4. The vertical dashed line shows the Fermi level, $E_F$. The density of states (DOS) describes the number of electronic states per unit energy per unit volume, which is available to be occupied. The finite values of TDOS at the Fermi level indicate that BaGa$_2$ is metallic. The value of TDOS at the $E_F$ is 2.20 states per eV per unit cell. At the Fermi energy level, the total DOS of BaGa$_2$ mainly comes from the contributions of Ba-4$d$ states and Ga-4$p$ states with some contributions of Ga-4$s$ states, as shown in the PDOS plots of Fig. 4.

The computed PDOS can be divided into three energy regions: the lowest energy region is due to the Ga-4$s$ electronic states; the region from -4 eV to 5 eV are coming from the Ba-4$p$ and Ga-4$d$ electronic states; the region above this is mainly due to the Ga-4$p$ and Ga-4$s$ electronic orbitals. The peaks in the TDOS close to the Fermi level controls the optical properties and electron scattering phenomena to a large extent. Strong overlap between orbitals in energy (e.g., Ba-4$d$ and Ga-4$p$) leads to hybridization and formation of covalent bonding between the atomic species. The presence of a pseudogap to the left of $E_F$ in the TDOS is a sign of significant structural stability of the compound [73].

The electron-electron interaction parameter of materials, also known as the Coulomb pseudopotential, can be estimated using the following relation [74]:

$$\mu^* = \frac{0.26 N(E_F)}{1 + N(E_F)} \qquad (29)$$

The total density of states at the Fermi level for BaGa$_2$ is 2.23 states/eV.unit cell. The electron-electron interaction parameter of BaGa$_2$ is therefore found to be 0.18. The transition temperature, $T_c$, of superconducting materials is reduced due to repulsive Coulomb pseudopotential [74 – 76].

### 3.5.3. Charge density distribution

We have calculated charge density distribution to explain the transfer of charge and visualize the nature of bonding among the atoms of BaGa$_2$. The electronic charge density distribution of BaGa$_2$ in different planes is displayed in Figure 5. The color scale on the right hand side of charge density maps displays the total electron density (blue and red color indicates high and low charge (electron) density, respectively). The accumulation of charges between two atoms indicates the covalent bonds, whereas the balancing of positive or negative charge at the atomic position indicates ionic bonding. Uniform charge smearing, on the other hand, illustrates metallic bonding. Maximum electron density is observed around Ga atoms as compared to Ba atoms. This is also consistent with the bond population analysis. This indicates the presence of ionic bonding between Ba and Ga atoms.

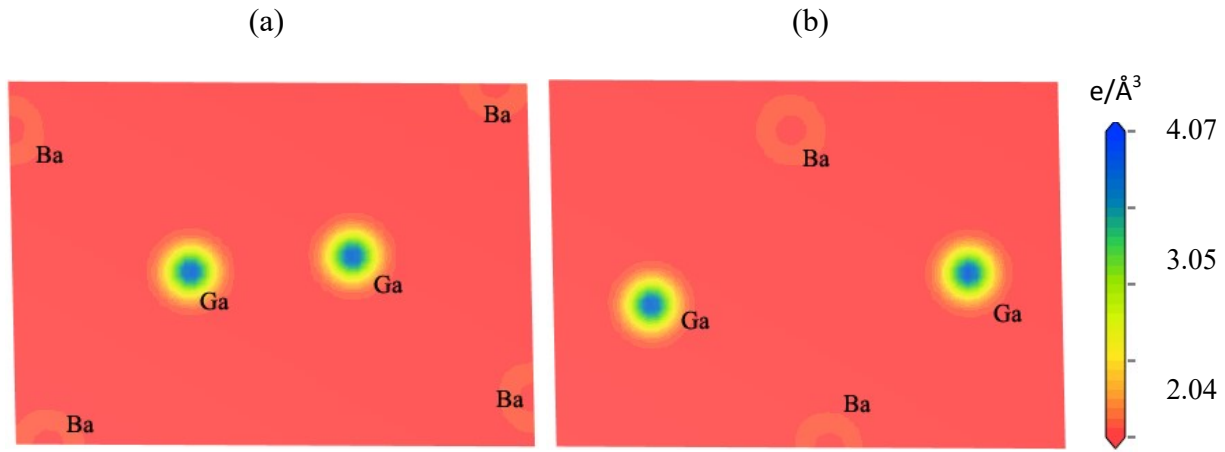

**Figure 5.** The electronic charge density distribution map (e/Å$^3$) of BaGa$_2$ in (a) (111) plane and (b) (001) plane.

### 3.5.4. Fermi surface

The study of Fermi surface of metallic materials is important since it gives us the idea of the behavior of occupied and unoccupied electronic states at low temperatures. A number of characteristics of a material, such as electronic, optical, thermal and magnetic properties are strongly dependent on Fermi surface topology. Superconductivity results from instability of the Fermi surface. The Fermi surface topology of the BaGa$_2$ has shown in Figs. 6 (a, b and c). Fermi surfaces of BaGa$_2$ are constructed form the bands crossing Fermi level with band numbers 17, 18 and 27 (Fig. 3). The Fermi surfaces show only electron-like sheets. For both bands 17 and 18, cylindrically co-axial electron-like sheets appear along the G-A path. For band 19, there are extended electron-like sheets close to the zone boundaries. These sheets are fairly dispersive in the *ab*-plane.

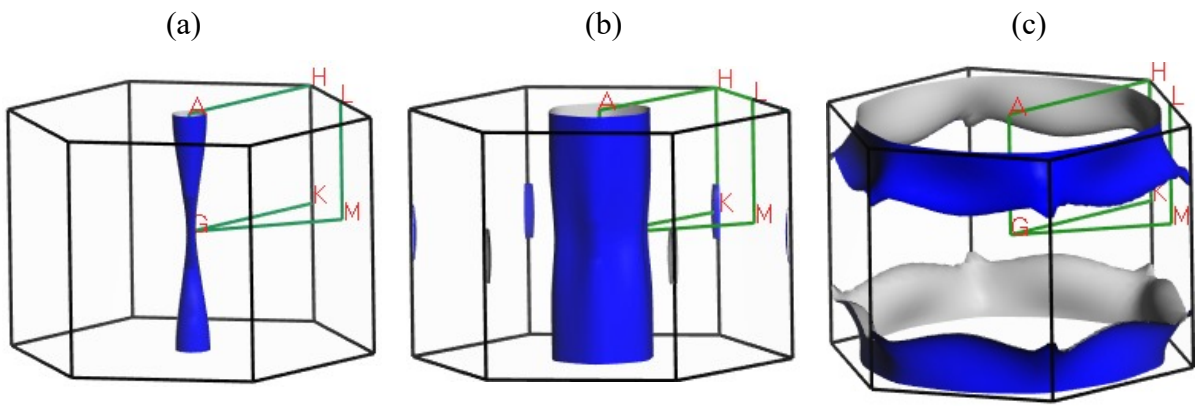

**Figure 6.** Fermi surface for the band numbers (a) 17 (b) 18 and (c) 19 of BaGa$_2$, respectively.

### 3.6. Phonon dispersion

Many properties of a material can be directly or indirectly determined from the phonon dispersion spectra and phonon DOS [77]. The phonon dispersion spectra (PDS) of a material contain valuable information regarding dynamical stability, phase transition and vibrational contributions in properties such as thermal expansion, Helmholtz free energy, and heat capacity [78]. The electron–phonon interaction function is directly related to the phonon density of states (DOS). The calculated PDS of BaGa$_2$ along the high symmetry directions of the crystal Brillouin zone (BZ) along with the total phonon density of states (PHDOS) in the ground state is depicted in Figure 7. The density functional perturbation theory (DFPT) finite displacement method (FDM) [15, 16] is employed to calculate the phonon characteristics which are shown in Figure 7. For application of a material under a time-varying mechanical stress, the dynamical stability of a material is an important criterion. The existence of the positive phonon frequency over the whole BZ confirms the dynamical stability of the compound. The highest vibrational frequency occurs around the *L*-point of the BZ with 5.88 cm$^{-1}$. This is an optical mode originating from the vibration of Ga atom.

For better understanding of the lattice dynamics, we have displayed total and atomic partial PHDOS of BaGa$_2$ in Figure 7, along with PDC to identify the bands and their corresponding density of states. The prominent peaks in the PHDOS are observed around 2.5 THz frequency. It is seen that the high-frequency phonon modes (> 5 THz) mainly result from the Ga atoms, whereas the low-frequency phonon modes (1 – 3 THz) come from both Ba and Ga atoms. The flatness of the bands produces the peaks in the PHDOS and highly dispersive bands decrease the heights of peaks in the total PHDOS.

The computed PHDOS can be divided into two regions encompassing acoustic modes and optical modes. The optical behavior strongly depends on the optical branches that are situated at the top in the phonon dispersive curve. It is well known that the total number of phonon branches is given by three times the number of atoms in a unit cell. Since each unit cell of BaGa$_2$ has 3 atoms, (Figure 1) this system will give rise to 9 normal lattice vibration modes from phonon dispersion curves, including 3 acoustical modes and 6 optical modes. The lower branches correspond to the acoustic modes (blue) and the upper branches with frequencies greater than 2 THz correspond to the optical modes (red). There is an overlap between the upper and lower branches of phonon dispersion curves; hence no phononic band gap exists between acoustic and optical modes in the PHDOS for BaGa$_2$. The existence of zero phonon frequency of the acoustic modes at the *Γ*-point is another indication of dynamical stability of the BaGa$_2$. It is seen from the PHDOS curve that both Ba and Ga atoms contribute to the acoustic and lower optical phonon modes. But the upper optical modes only originate from the vibrations due to lighter Ga atoms. The coherent vibrations of atoms in a lattice outside their equilibrium position cause acoustic phonon. On the other hand, when an atom moves to the left and its neighbor to the right, the optical phonon is originated due to the out of phase oscillations of the atom in a lattice. The optical properties of a crystal are mostly controlled by optical phonon.

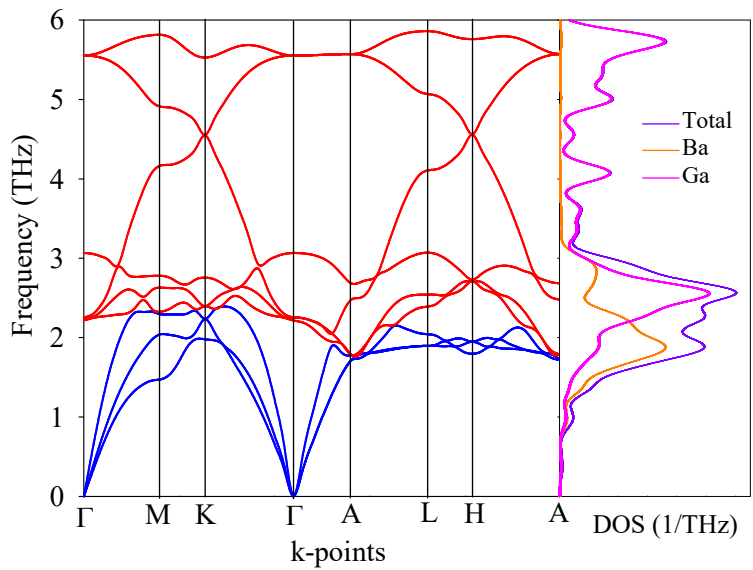

**Figure 7.** The phonon dispersion curve (PDC) and phonon density of states (PHDOS) of BaGa$_2$.

## 3.7. Bond population analysis

More insightful information about the chemical bonding (ionic, covalent and metallic) in solids can be obtained from the Mulliken population analysis (MPA) [27]. The calculations show that the total charge for Ga atoms is much larger than that for Ba atoms. This mainly comes from $3d$ states of Ga atom. The atomic charges of Ba and Ga in BaGa$_2$ are 0.14 and -0.07 electron, respectively (Table 9). Both of these values are deviated from their formal charge expected for purely ionic states (Ba: +2 and Ga: +3). Thus the electrons are partially transferred from Ba to Ga atoms. This deviation reflects the presence of some covalent bonds between Ba and Ga atoms. Since electrons are transferred from Ba to Ga atoms, ionic contributions are also present in BaGa$_2$. Thus the bonding behaviors are the combination of ionic and covalent bonds. The effective valence charge (EVC) for all atoms is defined as the difference between formal ionic charge and Mulliken charge within a crystal [79]. The zero (nonzero) value of EVC represents the existence of chemical bond as an ideal ionic bond (covalent bond) in the compounds and the higher values signifies the higher level of covalency in the bonding. The calculated effective ionic valences of BaGa$_2$ are disclosed in Table 9. It shows that the effective valences for both Ba and Ga are nonzero. This also indicates the presence of covalent in BaGa$_2$.

Despite the popularity of Mulliken bond population analysis, sometimes it gives results in contradiction to chemical intuition. This might be because of its strong atomic basis set dependency and simple basis sets lead to quite different deformation densities. To get more reliable result, Hirshfeld population analysis (HPA) [80] is employed. This approach does not require a reference to basis set or their respective location. Therefore, we have calculated Hirshfeld charge of BaGa$_2$ using the HPA. Table 9 shows comparison between Mulliken and Hirshfeld charge. Like Mulliken charge, Hirshfeld charge also predicts that electrons are transferred from Ba to Ga atoms. We have also estimated effective valence charges (EVC) of BaGa$_2$ from the Hirshfeld charge. Both the effective valence charges predict almost same level of covalency in BaGa$_2$.

**Table 9.** Charge spilling parameter (%), orbital charges (electron), atomic Mulliken charges (electron), formal ionic charge, EVC (electron) and Hirshfeld charge (electron) of BaGa$_2$ compound.

| Compound | Charge spilling | Species | Mulliken atomic population | | | | Mulliken charge | Formal ionic charge | EVC | Hirshfeld charge | EVC |
|---|---|---|---|---|---|---|---|---|---|---|---|
| | | | s | p | d | total | | | | | |
| BaGa$_2$ | 0.46 | Ba | 2.79 | 6.04 | 1.02 | 9.86 | 0.14 | +2 | 1.86 | 0.11 | 1.89 |
| | | Ga | 0.85 | 2.22 | 10.00 | 13.07 | -0.07 | +3 | 2.93 | -0.06 | 2.87 |

### 3.8. Optical properties

A material's response to incident electromagnetic radiation is related to its optical properties. The response to visible light is especially important from the view of optoelectronic applications. The response to incident radiation can be completely determined by various energy/frequency dependent parameters, namely the dielectric function, refractive index, loss function, optical conductivity, reflectivity and absorption coefficient. To explore the possible anisotropy in optical properties, we have calculated optical parameters of BaGa$_2$ for photon energies up to 15 eV for [100] and [001] polarization directions of the incident electric field. The optical properties are controlled by the electronic band structure and energy dependent density of states feature.

The frequency dependent real [$\varepsilon_1(\omega)$] and imaginary [$\varepsilon_2(\omega)$] parts of the dielectric function up to 15 eV are shown in Fig. 8(a). The real part of this function $\varepsilon_1(\omega)$ determines the polarizability of the compound, whereas the imaginary part is directly related to the band structure of the material and describes its absorptive behavior. Both intra-band and inter-band photon induced transitions contribute to $\varepsilon(\omega)$. Intra-band transitions play an important role at low energy in metals, whereas the inter-band term depends strongly upon the details of the electronic band structure [81] and peaks in the DOS. For example, the peaks in $\varepsilon_2(\omega)$ at ~2 eV and ~4.5 eV originate from the electronic transitions between the Ga-4$p$ and Ba-4$d$ states (Fig. 4). The low energy peak in the $\varepsilon_1(\omega)$ is characteristic of metallic systems. It is worth noting that the $\varepsilon_2(\omega)$ reaches zero from above at around 9.35 eV in the ultraviolet energy region, which indicates that BaGa$_2$ becomes fairly transparent above ~10 eV where optical absorption falls drastically. There is significant optical anisotropy in $\varepsilon_1(\omega)$ and $\varepsilon_2(\omega)$ with respect to the polarization of the incident electromagnetic radiation.

The knowledge of the refractive index of a compound is important for its use in optical devices such as photonic crystals, waveguides etc. The refractive index of BaGa$_2$ is calculated using following equation: $N(\omega) = n(\omega) + ik(\omega)$, where $k(\omega)$ is extinction coefficient. The refractive index ($n$) and the extinction coefficient ($k$) explain the phase velocity and the amount of absorption loss, respectively, when the electromagnetic wave passes through the material. The extinction coefficient ($k$) is a crucial parameter for photoelectronic device. Fig. 8(b) displays the variation of the refractive index for $BaGa_2$ as a function of incident photon energy. The calculated value of the static refractive index $n(0)$ is found to be at 8.18 and 4.61 along [100] and [001] directions, respectively. These values are quite high and the compound under study might be suitable for enhancing the visual aspects of electronic displays like LCDs, OLEDs, and quantum dot (QDLED) televisions. The peaks in $k(\omega)$ correspond closely to the spectrum of $\varepsilon_2(\omega)$ as expected.

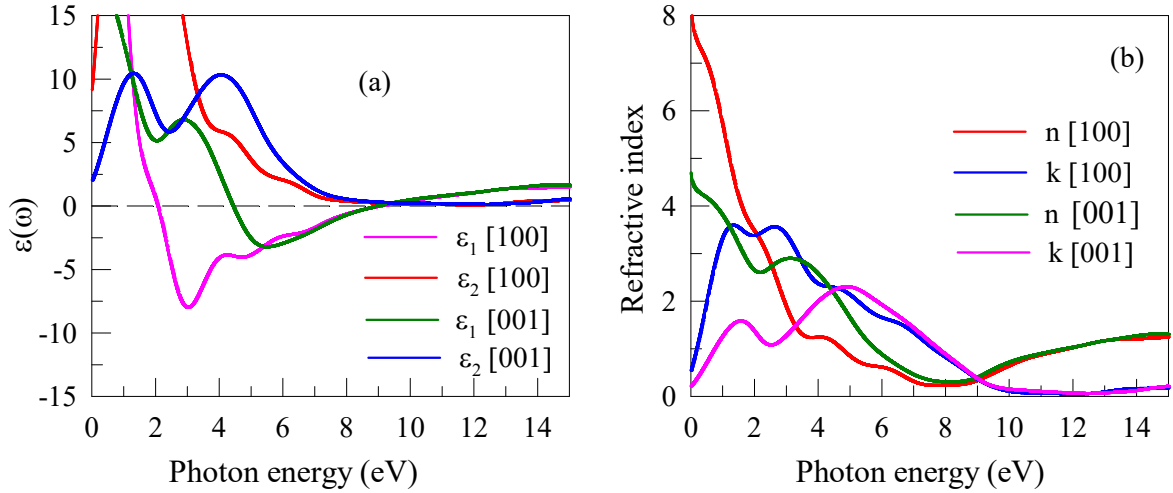

**Figure 8.** Photon energy dependence of (a) real part, $\varepsilon_1(\omega)$ and imaginary part, $\varepsilon_2(\omega)$ of the dielectric functions and (b) refractive index for $BaGa_2$.

The absorption coefficient ($\alpha$) provides information regarding the ability of a material to absorb incident radiation. Fig. 9(a) illustrates the absorption spectra of $BaGa_2$. The nonzero value of $\alpha$ in for both directions of polarization at zero photon energy is due to its metallic nature of $BaG_2$, which is consistent with the band structure and DOS calculations. The maximum absorption appears around 1.73 eV and 5.4 eV along [100] and [001] directions, respectively. In this sense, there is significant optical anisotropy in absorption characteristics. The real part of the photoconductivity ($\sigma$) spectrum of $BaGa_2$ is shown in Fig. 9(b). The nonzero photoconductivity at zero photon energy also indicates that $BaGa_2$ has no band gap, which is clearly seen in the electronic band structure calculations (Fig. 3). The maximum photoconductivity for this compound is found at 2.42 eV and 4.5 eV for [100] and [001] directions, respectively. Generally, the intraband contribution to the optical properties mainly dominated in the low energy infrared part of the spectra. On the other hand, in the high energy part of the absorption and conductivity spectra peaks may arise due to the interband transition. It is also apparent from Fig. 9a and b that the variation of the conductivity spectra is qualitatively similar to the absorption spectra. $BaG_2$

absorbs visible and near-ultraviolet radiation effectively for [100] electric field polarization. For the [001] polarization, absorption is strong in the energy range 5 – 7 eV in the ultraviolet region.

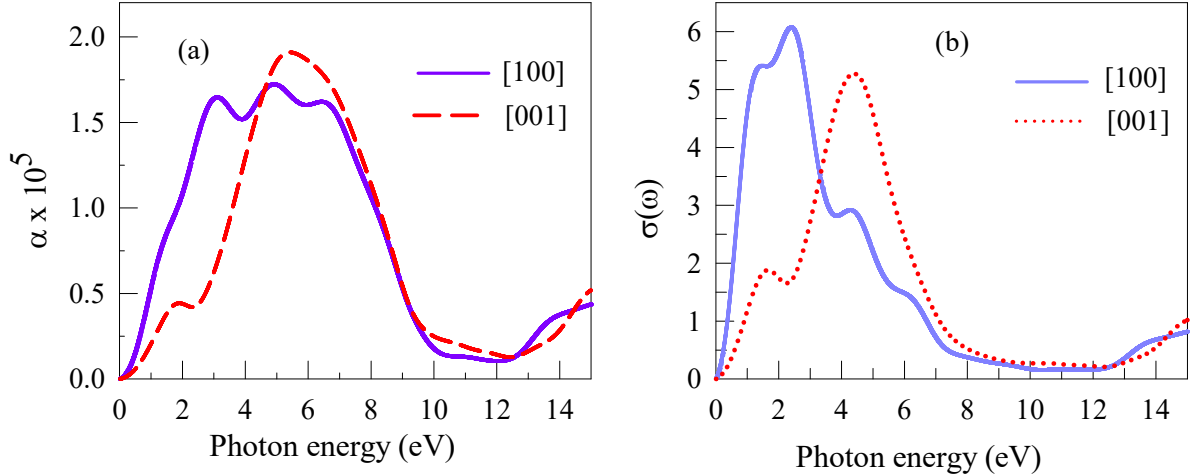

**Figure 9.** Photon energy dependence of (a) absorption coefficient, $\alpha(\omega)$ and (b) photoconductivity, $\sigma(\omega)$ for $BaGa_2$.

Fig. 10(a) exhibits the reflectivity spectrum of $BaGa_2$. The reflectivity is higher for [100] polarization than for [001]. $R(\omega)$ peaks at ~7.5 eV in the ultraviolet. $R(\omega)$ is almost nonselective and remains above 50% in the energy range from 0 to 8.0 eV when the electric field is polarized along [100] direction. This implies that $BaG_2$ can be used as an efficient reflector of solar radiation. The electron energy loss spectrum ($L$) of $BaGa_2$ is shown in Fig. 10(b). The energy loss function is a significant parameter to reveal the energy loss of a fast electron when it traverses in a material. In the loss function plot, the peaks are involved with the plasma resonance and the associated frequency is called the plasma frequency ($\omega_p$) [82, 83]. In addition, the sharp peaks of $L(\omega)$ also correspond to the trailing edges in the reflectivity and absorption spectra [84, 85]. For instance, the peak of $L(\omega)$ is at about 9.0 eV corresponding the abrupt reduction of $R(\omega)$ and $\alpha(\omega)$. The highest peak is found at about 9.06 eV for both the polarization directions, which discloses the plasma frequency of $BaGa_2$. When the incident light frequency is higher than the plasma frequency, the material becomes transparent and exhibits optical features similar to an insulator. Above plasma energy, $\varepsilon_2(\omega)$ approaches zero, signifying negligible loss of electromagnetic energy as it passes through the material.

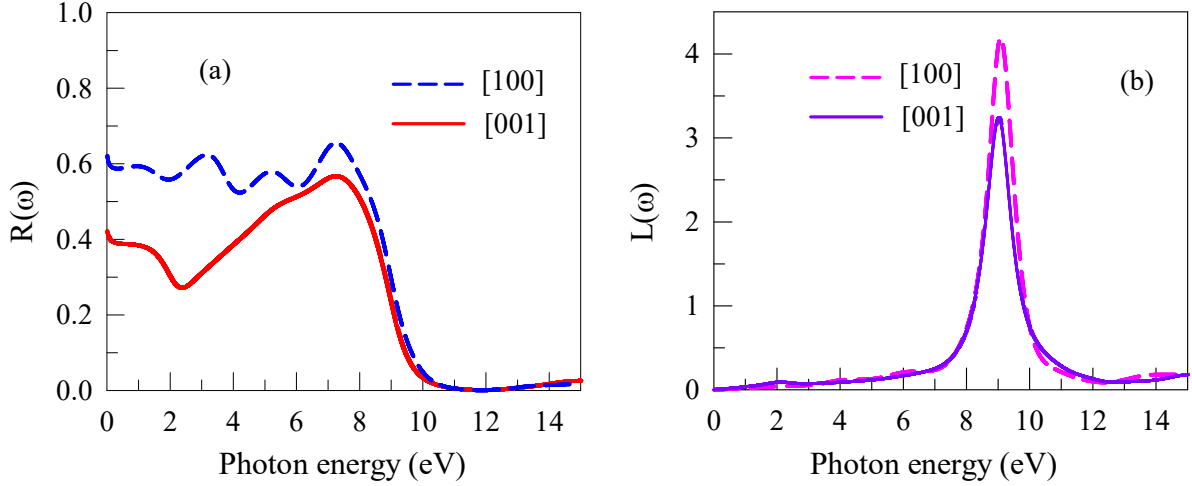

**Figure 10.** Photon energy dependence of (a) reflection coefficient, $R(\omega)$ and (b) loss function, $L(\omega)$ for $BaGa_2$.

### 3.9. Thermophysical properties
#### 3.9.1. Debye temperature

The study of Debye temperature ($\theta_D$) of a solid is important to comprehend many thermophysical properties such as bonding forces, the energy of formation of the vacancies, melting temperature, thermal conductivity, phonon dynamics, specific heat, superconductivity etc of solids. Generally speaking, the temperature at which the wavelength of phonons in a material corresponds roughly to the interatomic spacing is defined as the Debye temperature. The highest-frequency mode (and hence all the possible modes) of vibration is excited at $\theta_D$. This temperature is used to make a separation between high- and low-temperature regions for a solid. The Debye temperature also defines a border line to distinguish the classical and quantum behavior of phonons. When $T > \theta_D$, all modes of vibrations possesses same energy which is $k_BT$. But the higher frequency modes are considered to be absent at $T < \theta_D$ [86]. There are several methods for estimating $\theta_D$. At low temperatures the vibrational excitations are merely due to acoustic modes. Hence, at low temperatures the Debye temperature calculated from elastic constants is the same as that determined from the specific heat measurements. Here, we have calculated the Debye temperature of $BaGa_2$ from the elastic constants, which is one of the standard methods, using the following expression [68, 87]:

$$\theta_D = \frac{h}{k_B}\left(\frac{3n}{4\pi V_0}\right)^{1/3} v_a \tag{30}$$

where, $h$ is the Planck's constant, $k_B$ is the Boltzmann's constant, $V_0$ is the volume of unit cell and $n$ is the number of atoms in the compound.

The calculated value of the Debye temperature of BaGa$_2$ is displayed in Table 10. Debye temperature of BaGa$_2$ is obtained 203.99 K. Debye temperature of BaGa$_2$ is low which implies that the bonding strength among the atoms are not very strong and the compound under investigation is soft in nature.

### 3.9.2. Lattice thermal conductivity

For high temperature applications of materials, the lattice thermal conductivity is a crucial property to study. The lattice thermal conductivity ($k_{ph}$) of a solid at different temperature determines the amount of heat energy carried by lattice vibration. The $k_{ph}$ as a function of temperature is estimated from a formula derived by Slack [88 – 90] as follows:

$$k_{ph} = A(\gamma) \frac{M_{av}\theta_D^3 \delta}{\gamma^2 n^{2/3} T} \quad (31)$$

Here, $M_{av}$ is the average atomic mass per atom in a compound, $\theta_D$ is the Debye temperature, $\delta$ is the cubic root of average atomic volume, $n$ is the total number of atoms in the unit cell, $T$ is the absolute temperature and $\gamma$ is the Grüneisen parameter and $A(\gamma)$, which is a function of $\gamma$, is estimated from following equation (in W-mol/kg/m$^2$/K$^3$) [91]:

$$A(\gamma) = \frac{4.85628 \times 10^7}{2\left(1 - \frac{0.514}{\gamma} + \frac{0.228}{\gamma^2}\right)} \quad (32)$$

The calculated value of $k_{ph}$ for BaGa$_2$ is given in Table 10. According to Callaway–Debye theory [92], the low-temperature lattice thermal conductivity is proportional to Debye temperature $\theta_D$. A solid with a higher $\theta_D$ has a larger lattice thermal conductivity. Also, the lattice thermal conductivity and Young's modulus of a material are related as: $K_L \sim \sqrt{Y}$ [93].

### 3.9.3. Melting temperature

The melting temperature ($T_m$) is a parameter of interest which limits the temperature range of application of a material. A material with higher melting temperature implies that it will exhibit stronger atomic interaction, higher cohesive energy, higher bonding energy and lower thermal expansion [64, 94]. At temperatures below $T_m$, solids can be used continuously without oxidation, chemical change, and excessive distortion causing mechanical trouble. The melting temperature $T_m$ of a material can be obtained with the aid of elastic constants by the following empirical relation [95]:

$$T_m = 354K + (4.5K/GPa)\left(\frac{2C_{11} + C_{33}}{3}\right) \pm 300K \tag{33}$$

The estimated value of melting temperature of $BaGa_2$ is listed in Table 10. Generally, the melting temperature of a solid is closely associated with the presence of strong covalent bonding in the solid. The melting temperature of $BaGa_2$ is 640.76 ± 300 K. The melting temperature $T_m$ of a material is directly related its bonding strength.

### 3.9.4. Thermal expansion coefficient

Thermal expansion coefficient ($\alpha_T$) of a material is one of the intrinsic thermal properties. The thermal conductivity, specific heat, temperature variation of the energy band gap and electron effective mass of a solid are also related to its thermal expansion. It is also important for epitaxial growth of crystals and the reduction of harmful effects during its use in the electronic and spintronic devices. The thermal expansion coefficient of a material can be calculated from the following expression [64, 69]:

$$\alpha_T = \frac{1.6 \times 10^{-3}}{G} \tag{34}$$

where, $G$ represents the isothermal shear modulus (in GPa). The thermal expansion of solids is inversely related to their melting temperature: $\alpha \approx 0.02/T_m$ [69, 95]. The thermal expansion coefficient of $BaGa_2$ is disclosed in Table 10.

### 3.9.5. Dominant phonon mode and heat capacity

The long-wavelength phonons have a significant contribution to the thermal and electrical transport. The dominant phonon wavelength, $\lambda_{dom}$, is defined as the wavelength at which the phonon distribution function is peaked. This parameter depends on temperature and decreases with the increase of temperature. The wavelength of the dominant phonon for $BaGa_2$ at 300 K has been estimated from following expression [95]:

$$\lambda_{dom} = \frac{12.566 v_a}{T} \times 10^{-12} \tag{35}$$

where, $v_a$ is the average sound velocity in $ms^{-1}$, $T$ is the temperature in degree Kelvin. Materials with higher average sound velocity, higher shear modulus, lower density exhibits higher dominant phonon wavelength [95]. The estimated value of $\lambda_{dom}$ of $BaGa_2$ is listed in Table 10.

The heat capacity ($C_p$) is a fundamental thermodynamic property of a material. This parameter is crucial for many thermodynamic calculations and engineering applications (material designing process). Materials with higher heat capacity possess higher thermal conductivity and lower thermal diffusivity. The change in thermal energy per unit volume in a material per Kelvin change in temperature explains heat capacity per unit volume ($\rho C_P$). The heat capacity per unit volume can be evaluated from following expression [64, 69]:

$$\rho C_P = \frac{3k_B}{\Omega} \quad (36)$$

where, $N = 1/\Omega$ defines the number of atoms per unit volume. The heat capacity per unit volume of BaGa$_2$ is shown in Table 10.

**Table 10.** The Debye temperature $\Theta_D$ (K), lattice thermal conductivity $k_{ph}$ (W/m-K) at 300 K, melting temperature $T_m$ (K), thermal expansion coefficient $\alpha_T$ (K$^{-1}$), wavelength of dominant phonon $\lambda_{dom}$ (m) at 300 K, and heat capacity per unit volume $\rho C_P$ (J/m$^3$-K) of BaGa$_2$.

| Compound | $\Theta_D$ | $k_{ph}$ | $T_m$ | $\alpha$ ($\times 10^{-5}$) | $\lambda_{dom}$ ($\times 10^{-12}$) | $\rho C_P$ ($\times 10^6$) | Ref. |
|---|---|---|---|---|---|---|---|
| BaGa$_2$ | 203.99 | 4.78 | 640.76 | 8.03 | 90.55 | 1.40 | This work |
| | 224.80 | - | - | - | - | - | [3]$^{Theo.}$ |

### 3.9.6. Minimum thermal conductivity and its anisotropy

The minimum thermal conductivity is an extreme limit of a basic thermal property. The theoretical minimum thermal conductivity ($K_{min}$) is defined as the lowest limit of thermal conductivity of a crystal above Debye temperature. Hence, it is important to calculate $K_{min}$ of a crystal for high temperature applications. According to Clarke model, the minimum thermal conductivity $k_{min}$ can be calculated from [95]:

$$k_{min} = k_B \upsilon_a (V_{atomic})^{-2/3} \quad (37)$$

In this equation, $k_B$, $\upsilon_a$ and $V_{atomic}$ refer to the Boltzmann constant, average sound velocity and cell volume per atom, respectively. Materials with higher acoustic velocity, minimum phonon mean free path and Debye temperature have higher minimum thermal conductivity. The evaluated value of isotropic minimum thermal conductivity for BaGa$_2$ is enlisted in Table 10.

Materials with anisotropic thermal conductivity have many applications, for instance for heat spreading in electronic and optical device technologies as well as in heat shields, thermoelectrics and thermal barrier coatings. Apart from layered composites, elastically anisotropic compounds offer the best opportunities for exhibiting intrinsic anisotropic thermal conductivity, since propagation of elastic wave dominates heat transfer. Thermal vibrations of atoms, movement of free electrons in metals, radiation are three different modes of transmission of heat through solids. To discuss the anisotropy of thermal conductivity, the minimum thermal conductivities of BaGa$_2$ along different directions are also investigated employing the Cahill's model [96]:

$$k_{min} = \frac{k_B}{2.48} n^{2/3} (\upsilon_l + \upsilon_{t1} + \upsilon_{t2}) \qquad (38)$$

and
$$n = N/V$$

where, $k_B$ is the Boltzmann constant, $n$ is the number of atoms per unit volume and $N$ is total number of atoms in the cell having a volume $V$.

**Table 11.** The number density of atoms per mole of the compound $n$ (m$^{-3}$), minimum thermal conductivity (W/m-K) of BaGa$_2$ along different directions evaluated by Cahill's and Clarke's method.

| Compound | $n$ (10$^{28}$) | [100]$k_{calc.}^{min}$ | [001]$k_{calc.}^{min}$ | $k_{min}$ | |
|---|---|---|---|---|---|
| | | | | Cahill | Clarke |
| BaGa$_2$ | 3.39 | 0.443 | 0.447 | 0.441 | 0.320 |

The minimum thermal conductivity of BaGa$_2$ in [100] and [001] directions are summarized in Table 10. The compound shows negligible anisotropy and low $K_{min}$. We have also calculated the isotropic minimum thermal conductivity using Cahill's method [96]. The estimated results are presented in Table 10. The isotropic minimum thermal conductivity using Cahill's and Clarke's methods are 0.441 and 0.320, respectively. Clearly, Cahill's model predicts slightly larger $K_{min}$ compared with Clarke's model. This has been also observed for other compounds [64, 65, 97]. This probably arises because the Cahill's model considered the atomic number density and the whole phonon spectrum, while the Clarke's model does not take the contributions of the optical phonons in consideration [98]. Therefore, one expects that the result obtained by Cahill's model should be closer to the real values than that obtained by Clarke's model.

## 4. Conclusions

In this paper we have investigated a large number of hitherto unexplored physical properties of binary topological semimetal BaGa$_2$ for the first time. The compound is found to be mechanically stable from both elastic and phonon dispersion calculations. The compound is fairly machinable, with weakly brittle features. The bondings between atoms have mixed character with ionic and covalent contributions. The hardness of BaGa$_2$ is low. The electronic band structure, density of states and optical calculations reveal semimetallic characteristics with clear topological signatures when SOC is included. Quasi-2D Dirac cone found at the $K$-point of the Brillouin zone touching the Fermi level suggests that charge carries have high mobility in these bands. The calculated value of the Coulomb pseudopotential is large [99] and might be responsible partly for the low superconducting transition temperature found in BaG$_2$. The main contribution to the $N(E_F)$ comes from the Ba 4$d$ and Ga 4$p$ electronic states. Electronic dispersion shows anisotropy. The charge density distribution, MPA and HPA results are consistent with each other. The value of the universal log-Euclidean index ($A^L$) implies that the bonding strength in BaGa$_2$ is anisotropic along different directions within the crystal.

The calculated value of Debye temperature is low and found to be in good agreement with previously estimated value [3]. It is instructive to note that calculated Debye temperature of BaGa$_2$ is comparable to some other binary intermetallic compounds [100]. The melting temperature and minimum thermal conductivity of BaGa$_2$ are also low. All the estimated thermophysical properties exhibit good correspondence with each other.

The optical parameters of BaGa$_2$ have been explored in details for the first time. BaGa$_2$ possesses high reflectivity and absorption coefficient over extended range of photon energy. The material also has a large static refractive index. Nearly nonselective and high reflectivity of BaG$_2$ indicates that this compound has potential to be used as a coating material to reduce solar heating. The static and low energy refractive index is also quite high making BaG$_2$ a potential material for optoelectronic display devices. The optical parameters of BaG$_2$ possess notable anisotropy with respect to the polarization of the incident electromagnetic field. The optical constant spectra agree very well with the calculated electronic band structure.

In summary, we have investigated the elastic, mechanical, bonding, acoustic, thermal electronic, phonon dispersion and optical properties of BaGa$_2$ in details in this paper. The compound under study possesses several attractive mechanical, thermal and optoelectronic features which will be suitable for engineering and device applications. We anticipate that the results obtained in this paper will stimulate researchers to investigate BaGa$_2$ in greater details in future, both theoretically and experimentally.

**Acknowledgements**
S. H. N. acknowledges the research grant (1151/5/52/RU/Science-07/19-20) from the Faculty of Science, University of Rajshahi, Bangladesh, which partly supported this work.

**Data availability**
The data sets generated and/or analyzed in this study are available from the corresponding author on reasonable request.

## Author Contributions

## Additional Information
### Competing Interests